\documentclass[aps,prl,floatfix,twocolumn]{revtex4-2}

\usepackage{amsmath,amssymb}
\usepackage[pdftex]{hyperref,graphicx}
\hypersetup{colorlinks = true, urlcolor = blue, linkcolor = blue, citecolor = blue}

\usepackage{physics}
\usepackage{xcolor}
\usepackage{bm}

\begin{document}
	\title{Interaction-Driven Filling-Induced Metal-Insulator Transitions in 2D Moir\'e Lattices}
	
	\author{Haining Pan}
	\affiliation{Condensed Matter Theory Center and Joint Quantum Institute, Department of Physics, University of Maryland, College Park, Maryland 20742, USA}

	\author{Sankar Das Sarma}
	\affiliation{Condensed Matter Theory Center and Joint Quantum Institute, Department of Physics, University of Maryland, College Park, Maryland 20742, USA}
	
\begin{abstract}
	Using a realistic band structure for twisted WSe$_2 $ materials, we develop a theory for the interaction-driven correlated insulators to conducting metals transitions through the tuning of the filling factor around commensurate fractional fillings of the moir\'e unit cell in the 2D honeycomb lattice, focusing on the dominant half-filled Mott insulating state, which exists for both long- and short-range interactions.  We find metallic states slightly away from half-filling, as have recently been observed experimentally.  We discuss the stabilities and the magnetic properties of the resulting insulating and metallic phases, and comment on their experimental signatures. {We also discuss the nature of the correlated insulator states at the rational fractional fillings.}
\end{abstract}
\maketitle

The Fermi liquid {(FL)} theory is the most successful paradigm in condensed matter physics asserting that an interacting many-fermion system in dimensions higher than one (e.g., metals, normal He-3) has a {bijection} with the noninteracting Fermi gas. The low-energy quasiparticle excitations of the interacting Fermi system behave as almost noninteracting excitations of the Fermi gas with renormalized properties such as the effective mass.  A well-known simple exception to the {FL} paradigm was pointed out by Wigner rather early~\cite{wigner1934interaction} where he showed that strong long-range Coulomb interactions, would crystallize a continuum electron gas, creating a Wigner crystal (WC) of electrons, so that the Coulomb potential energy of the electrons is minimized instead of the kinetic energy as in the noninteracting or the weakly interacting system.  Obviously, the WC is not a {FL}. Later, Mott~\cite{mott1949basis} argued that interacting band electrons in a lattice would undergo correlation-driven metal to insulator transition for strong enough interactions. The Mott transition is adiabatically connected to the Wigner transition~\cite{kohn1967mott,vu2020collective}. The concept of Mott transition evolved over time eventually, becoming a sharply defined paradigm as the Mott-Hubbard (MH) metal-insulator transition {(MIT)}~\cite{hubbard1963electron}. The modern view of the MH transition~\cite{mott1968metalinsulator,gebhard1997mott,mott2004metalinsulator} involves the correlation-driven {MIT} at the half-filling of a narrow tight-binding band with the electrons being localized at lattice sites as local magnetic moments in an antiferromagnetic insulating (AFI) state.  Such an AFI, existing precisely at the half-filling of the original noninteracting band, is called a Mott insulator (MI), and it arises from the strong short-range interactions present in the Hubbard model preventing the double occupancy of lattice sites, thus creating a purely on-site interaction driven insulating state.  Such a MI is quite distinct from the WC in three specific ways: (1) it is independent of the electron density, and does not necessitate a low-density electron system as the WC requires; (2) it arises purely from short-range correlation effects in contrast to the WC arising from the long-range Coulomb interaction; (3) the MI happens precisely at the half-filling of the noninteracting band with the average inter-electron separation being equal to the effective lattice constant (i.e., one electron per unit cell).  Although both the WC and MI are interaction-driven insulators, the MI is ubiquitous in strongly correlated narrow band systems~\cite{imada1998metalinsulator} whereas the pure WC is rarely experimentally observed~\cite{grimes1979evidence}. The current work studies the interplay between WC and MI phenomena driving {(MIT)} in the context of actual experiments in twisted 2D moir\'e systems based on van der Waals materials~\cite{wang2020correlated,regan2020mott,zhang2020flat,tang2020simulation,xu2020correlated,huang2021correlated,jin2021stripe,cao2018unconventional,cao2018correlated,yankowitz2019tuning,lu2019superconductors}. In particular, our focus is on the physics of the filling close to half, where a metallic state could exist at small doping away from the half-filled MI state.  Such correlated metallic states {themselves are interesting} in addition to the considerable interest in the physics of {MIT}.

Recent experiments have identified correlated insulating states in twisted transition metal dichalcogenides (TTMD) based moir\'e 2D systems at filling factors $\nu$=1, 3/4, 2/3, 1/2, 1/3, 1/4, etc.~\cite{ huang2021correlated,jin2021stripe,xu2020correlated,zhang2020flat,tang2020simulation,regan2020mott,wang2020correlated}{, where $ \nu $ denotes the number of holes per moir\'e unit cell.}
In TTMD moir\'e systems, strong spin-orbit coupling produces doubly degenerate flat hole bands with narrow effective bandwidth leading to strong correlation effects as the kinetic energy is exponentially suppressed.  In addition to the filling factor $\nu$, the system has two other tuning parameters affecting correlation effects: the twist angle $\theta$ determining the moir\'e unit cell size and the effective dielectric constant $\epsilon$ determining the Coulomb coupling as defined by $ e^2/(\epsilon r) $ where $r$ is the interparticle distance.  $\theta$ defines the moir\'e band structure, and $\epsilon$ defines the Coulomb coupling strength whereas $\nu$ defines the band filling.

{We start with a MI at $ \nu $=1,} and investigate if correlated metallic states could exist in its neighborhood, as has recently been observed in transport measurements~\cite{wang2020correlated}.  We note that the phase diagram of the {TTMD} at fixed rational fillings has recently been calculated~\cite{pan2020quantum} and experimentally studied~\cite{jin2021stripe,xu2020correlated}. The insulating states arising at various fractions (e.g., $\nu=$1/3, 1/2, etc.) are neither strict MI nor strict WC. They are {best described} as `correlated insulators’ {(CIs)}.  They are Mott-like {because} they are commensurate with the moir\'e lattice and are thus connected to the band physics, but they are Wigner-like {because} their existence, except for the $\nu$=1 {(MI)}, depends on {long-ranged Coulomb potential.} These are all {CIs} specific to moir\'e systems, which are neither MI nor WC.

{The} realistic interacting tight-binding Hamiltonian for TTMD-based moir\'e systems {is}~\cite{pan2020band,pan2020quantum}
\begin{eqnarray}\label{eq:hubbard}
	H&=&\sum_{s}\sum_{i,j}^{} t_{s}\left(\bm{R}_i-\bm{R}_j\right) c_{i,s}^\dagger c_{j,s}\nonumber\\
	&+&\frac{1}{2}\sum_{s,s'}\sum_{i,j}U(\bm{R}_i-\bm{R}_j) c_{i,s}^\dagger c_{j,s'}^\dagger c_{j,s'} c_{i,s},
\end{eqnarray}
where {the hopping terms $t_s$ represent band structures (depending on $\theta$), and effective interparticle Coulomb interactions $ U $  represent the correlation effect (depending on $\epsilon$)}. Valley index $s$ , spin-up or down, is coupled with $+K$  or $-K$  valley, respectively, in the Brillouin zone~\cite{pan2020band}. {Both} $ t $ and $ U $ involve distant nearest neighbors ({i.e., our parametrization of Eq.~\eqref{eq:hubbard} includes hopping up to the third nearest neighbors and Coulomb coupling term $ U $ up to 1993 distant sites}), refer to Refs.~\onlinecite{pan2020band,pan2020quantum,wu2018hubbard,wu2019topological} for the motivation and derivation of Eq.~\eqref{eq:hubbard} as the basic description for the interacting moir\'e physics in TTMD systems.  Although the theory based on Eq.~\eqref{eq:hubbard} should apply to all TTMD systems, our specific numerical results are for the WSe${}_2$ based TTMD structures currently being studied at Columbia University~\cite{wang2020correlated}.  Details of the numerical model for Eq.~\eqref{eq:hubbard} {are available in Refs.}~\onlinecite{pan2020band,pan2020quantum}.  We emphasize that Eq.~\eqref{eq:hubbard} cannot be thought of as either a Hubbard (or extended Hubbard) model or a {WC} (or generalized WC) model. Equation~\eqref{eq:hubbard} is a semirealistic model for the actual interacting TTMD 2D moir\'e materials.

The 2D interacting problem in Eq.~\eqref{eq:hubbard} is well defined, once all the hopping terms $ t $ and interaction terms $ U $, along with the filling factor $\nu$, are known.  
Obviously, the problem is insoluble {exactly}: {the fermion sign problem and} the 2D nature of the system make {quantum Monte Carlo or exact diagonalization impossible}.
{When} the first term in Eq.~\eqref{eq:hubbard} is zero, the problem has an exact classical solution which is obtained by minimizing the Coulomb energy (i.e., the second term).  The exact classical solution for $ t $=0 depends on the precise {$\nu$} since the lattice symmetry of the classical state depends on {$\nu$}.  Our strategy, as explained in Refs.~\onlinecite{pan2020band,pan2020quantum}, is to use a self-consistent mean-field {(SCMF)} theory starting with the classical WC solution as the initial input to obtain the final ground state of Eq.~\eqref{eq:hubbard} in the presence of {the hopping term}.  This is a reasonable strategy to search for {CI} ground states at rational fillings in the presence of strong interactions. It is possible that our theory overestimates the importance of non-{FL} correlated insulating states over {FL} conducting metallic states, which is acceptable since the problem is {interesting} only because of the breakdown of {FL} theory in the interacting system leading to the insulating states, which are absent (except trivially at band filling, $\nu$=2) in the tight-binding problem without interactions. For small deviations in $ \nu $ around a rational filling, one can study the possible emergence of correlated metallic states by using a perturbation theory around the mean-field solution. {Our current work is qualitatively different from earlier work since we are dealing here with metallic ground states in contrast to insulating ground states in~\cite{pan2020band,pan2020quantum}}.
\begin{figure}[t]
	\centering
	\includegraphics[width=3.4in]{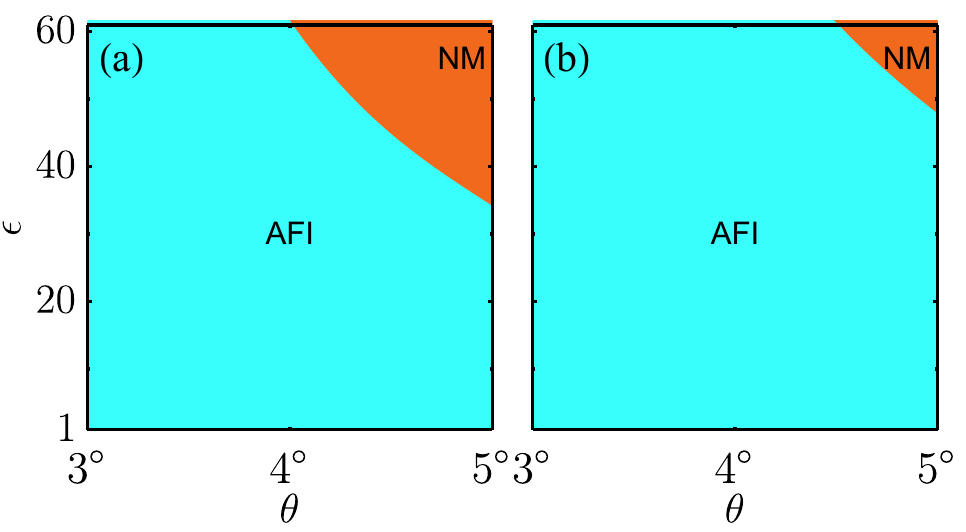}
	\caption{The calculated phase diagram in the $ \epsilon $-$ \theta $ space at $ \nu $=1 for (a) all distant neighbor Coulomb couplings, and (b) only the on-site $ U $.}
	\label{fig:fig1}
\end{figure}

In Fig.~\ref{fig:fig1}, we show the calculated quantum phase diagram for the $\nu$=1 half-filled state in the $\theta$-$\epsilon$ space for the full long-range interaction with all distant neighbor Coulomb couplings as well as the effective Hubbard model keeping only the on-site {$ U $}. The two phase diagrams are almost identical (except for some small quantitative differences), with an AFI state being the dominant phase except at large $\epsilon$ (i.e., weak interaction) and large $\theta$ (i.e., strong hopping) where the paramagnetic normal metallic (NM) {FL} shows up.  This $\nu$=1 AFI is the well-known MI phase. We emphasize that Fig.~\ref{fig:fig1}, manifesting a similar MI to NM transition for both long-range and on-site interaction models, serves as an important check on our theory. The fact that the theory reproduces the correct {MI} at $\nu$=1 irrespective of whether the interaction is long-ranged Coulomb or short-ranged Hubbard establishes the correct qualitative reliability of the theory.

Having established the model and its validity in predicting the correct {MI} at $\nu$=1, we now consider the main thrust of this Letter: filling-induced {MIT} in the flatband moir\'e system.  We must distinguish among three distinct types of MIT predicted by the solutions to Eq.~\eqref{eq:hubbard} in the moir\'e system.  First, there are two types of correlation-driven MITs for fixed $\nu$, tuned respectively by $\epsilon$ and $\theta$, as apparent in Fig.~\ref{fig:fig1}.  Increasing $\epsilon$ suppresses interaction and increasing $\theta$ increases the effective bandwidth, so tuning either $\epsilon$ or $\theta$ is an essentially equivalent way of changing the dimensionless interaction strength $U/t$, although, unlike in the simple Hubbard model, both $ U $ and $ t $ are represented by many effective {parameters in} Eq.~\eqref{eq:hubbard} instead of a single {parameter}. Similar correlation-tuned $\epsilon$-$\theta$  phase diagrams are provided in Refs.~\onlinecite{pan2020band,pan2020quantum} for $\nu$=3/4, 2/3, 1/2, 1/3, 1/4. We note that tuning the interaction strength \textit{in situ} at a fixed $\nu$ is a challenge experimentally since a typical sample has a fixed twist angle and {substrate}, and changing samples to change $\theta$ or $\epsilon$ may lead to other unknown modifications. {Actually}, $ \epsilon/\theta $-tuned MIT has not yet been observed in TTMD [or {twisted bilayer graphene (TBLG)}], where a system is experimentally found to be either insulating or metallic depending on {$\nu$} at low temperatures.  

More interesting and accessible is the third type of MIT, which is tuned by {$\nu$} at fixed $\epsilon$ or $\theta$. In the filling-tuned MIT, one dopes (or gates) the system away from a fixed $\nu$ where additional holes or electrons are created because of doping, and the system could undergo an insulator to metal transition solely because of doping (i.e., variation in $\nu$) itself without any explicit change in $\epsilon$ or $\theta$. Such $\nu$-tuned MIT has been reported in the WSe$_2$ TTMD structures, where our theory should be applicable~\cite{wang2020correlated}. Our theory is partially motivated by the recent experiments at Columbia University~\cite{wang2020correlated}, which is also recently observed in the twisted WSe$ _{2} $/WSe$ _{2} $~\cite{ghiotto2021quantum} and a similar heterostructure system MoTe$ _{2} $/WSe$ _{2} $~\cite{li2021continuous}.
\begin{figure}[t]
	\centering
	\includegraphics[width=3.4in]{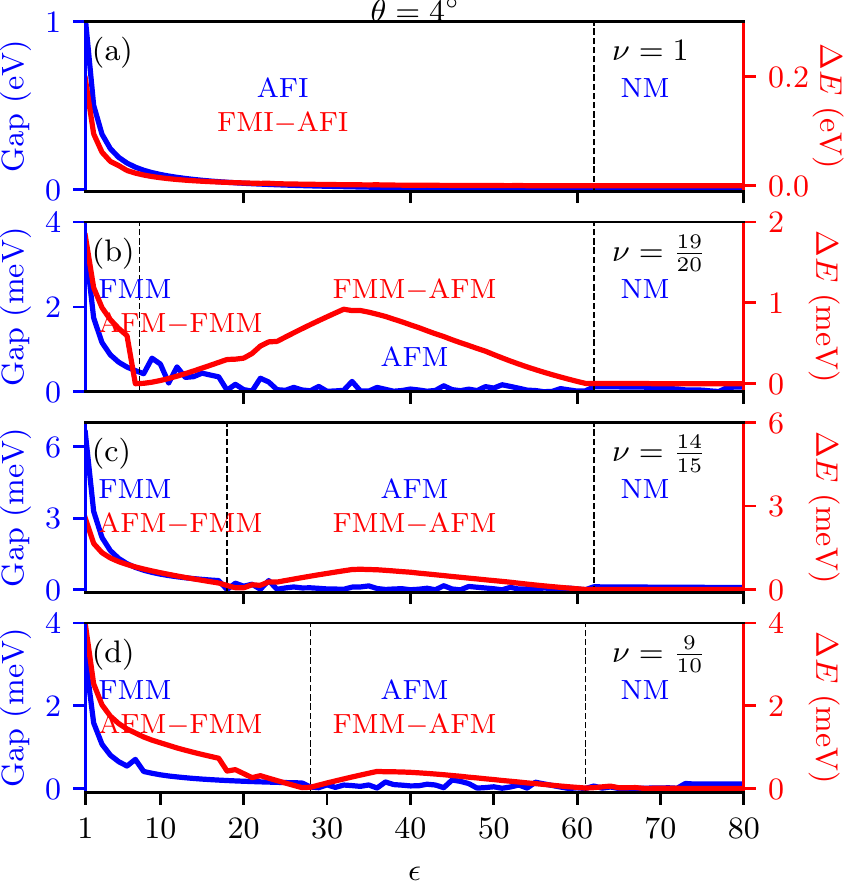}
	\caption{The calculated charge gap {(blue)} as a function of $ \epsilon $ for a fixed $ \theta $=4$^\circ$ at (a) $ \nu $=1; (b) $ \nu $=19/20; (c) $ \nu$=14/15; (d) $ \nu $=9/10. {The red line} is the absolute energy difference between two competing phases. The first order transition happens at the vertical dashed line with the name of phases labeled in blue. AFI, AFM, FMI, FMM, and NM denote respectively the antiferromagnetic insulator, antiferromagnetic metal, ferromagnetic insulator, ferromagnetic metal, and normal paramagnetic metal.}
	\label{fig:fig2}
\end{figure}

\begin{figure}[t]
	\centering
	\includegraphics[width=3.4in]{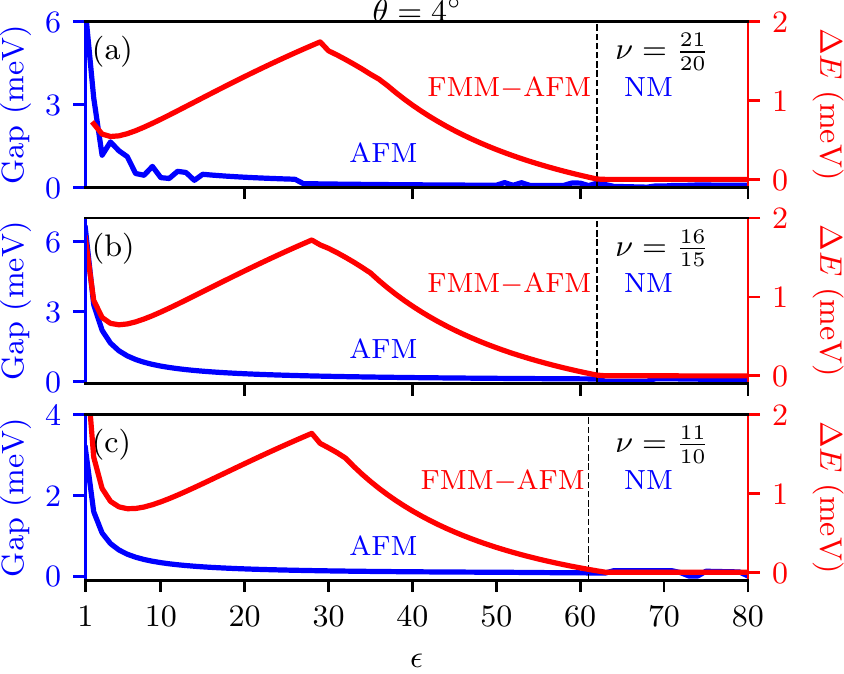}
	\caption{The calculated charge gap {(blue)} as a function of $ \epsilon $ for a fixed $ \theta$=4$^\circ $ at (a) $ \nu $=21/20; (b) $ \nu $=16/15; (c) $ \nu $=11/10. {The red line} is the absolute energy difference between closest competing ground states. Refer to Fig. 2 for notations.}
	\label{fig:fig3}
\end{figure}

In Figs.~\ref{fig:fig2}(b)-\ref{fig:fig2}(d), we {show} calculated charge (energy) gaps {at Fermi energy} as a function of $\epsilon$ for fixed {$\theta$ =4$ ^\circ $} for several values of $\nu$, close to but slightly below $\nu$=1.  We also show the same quantity for $\nu$=1 for the sake of comparison in Fig.~\ref{fig:fig2}(a). Our calculation is perturbative starting from the $ \nu=1 $ {MI} which we calculate nonperturbatively using {SCMF} theory. The experimentally relevant value of $\epsilon$ depending on the sample and experimental details is approximately between 5 and 30, {most likely 10--15}.  It is remarkable that although the $\nu$=1 state is a strong Mott AFI with a large gap for {reasonable} $\epsilon$, the situation is qualitatively different for $\nu$ just slightly below unity, where the ground state is a ferromagnetic metal (FMM) or antiferromagnetic metal (AFM) depending on  $ \epsilon $ in contrast to the antiferromagnetic MI at $\nu$=1. (The AFM phase {is} a spin density wave adiabatically connected to the N\'eel AFI at $ \nu $=1.) This is the filling-induced MIT, where the states at {$\nu$= 1, 2/3, 1/2, 1/3, etc.} are {CI}, but nearby states doped slightly away are metallic. The absolute energy difference between the two competing phases (AFM versus FMM) near $ \nu\sim$ 1 is {in} red {corresponding} to the right axis in Fig.~\ref{fig:fig2}.  We mention that the small gap {at} small $\epsilon$ (i.e., strong interactions) in the effectively metallic ground state {reflects} that the calculation is always done at a rational filling factor, not at a true incommensurate filling which would be the generic experimental situation just slightly away from half-filling. This numerical gap in the metallic phase is fictitious due to the finite discretization of the momentum space, which should vanish {as the mesh in the momentum space becomes finer, i.e., it is metallic in the thermodynamic limit.} {Note} that this gap is orders of magnitude smaller than the calculated gap at $\nu$=1 and the gap goes quickly to zero with increasing $\epsilon$.  
This ensures that our theory captures the correct physics of generic metallicity at incommensurate filling close to $\nu\sim$ 1{, at small $\epsilon$ (large interaction), which approaches the limit where Nagaoka-like ferromagnetism dominates}.  
{For $ \nu >$ 1, we find similar metallic state but without the Nagaoka ferromagnetism. This is because the filling denotes the doping of hole and it is equivalent to taking out the electrons if $\nu$ goes above 1. In the electron number basis, the hopping is negative, where the Nagaoka theorem requires a positive hopping, thus the Nagaoka ferromagnetism is not guaranteed at $\nu>1$ .}
Figure~\ref{fig:fig3} shows the representative theoretical results.
We note that the $ \nu>$ 1 manifests an {AFM} phase in contrast to {Nagaoka-like} FMM for $ \nu<$ 1.

{The first transport experiment on WSe${}_2$ already showed the existence of a strong insulating phase at $\nu$=1 and metallic phases for doping slightly away from half-filling~\cite{wang2020correlated}. So, our finding of a filling-tuned MIT in TTMD moir\'e system around $ \nu\sim $1 is consistent with experimental results~\cite{ghiotto2021quantum}.
We have also checked that similar metallic phases exist around the other TTMD {CIs} at fractional occupancy {e.g., $ \nu $=1/3}~\cite{MI_SM}.}

\begin{figure}[t]
	\centering
	\includegraphics[width=3.4in]{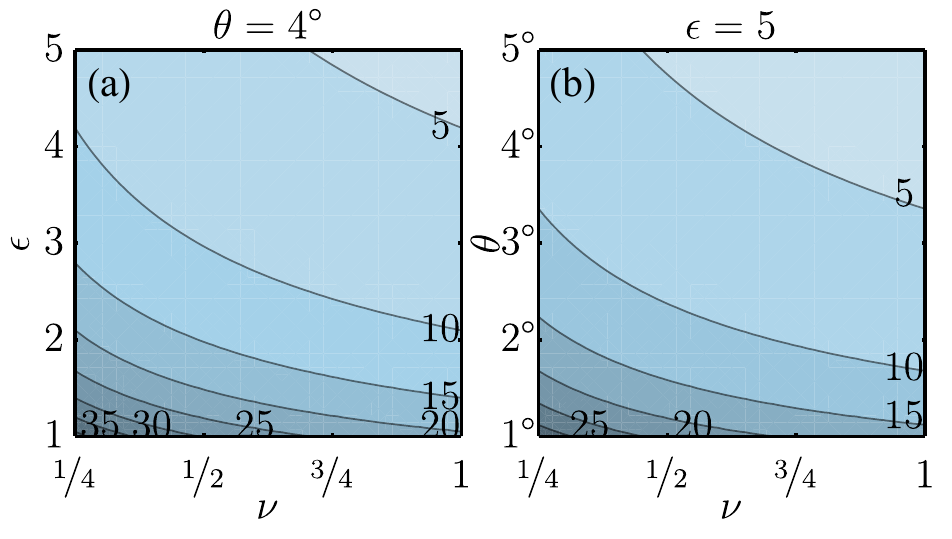}
	\caption{The dimensionless continuum Coulomb coupling $ r_s $ (a) at a fixed $ \theta$=4$ ^\circ $; (b) at a fixed $ \epsilon$=5. The lattice constant of the monolayer WSe$ _2 $ is 3.28\AA~\cite{kormanyos2015cdotp}, and effective mass $ m^*$=0.45$m_e $~\cite{fallahazad2016shubnikov} ($ m_e $ is the rest electron mass). The 2D quantum WC condition $ r_s>$ 30 is {unsatisfied} in the experimentally relevant regime.}
	\label{fig:fig4}
\end{figure}

{Having established that a metallic state exists generically away from half-filling (and other rational) filling in the system, we now briefly discuss the nature of the correlated insulating states, which should all be metallic states in the noninteracting band picture because they represent fractionally filled single-particle energy bands of the TTMD system.} We have already emphasized that the interaction driven insulating phase {at fractional rational fillings}~\cite{pan2020band,pan2020quantum} arising in moir\'e TTMD materials should be thought of simply as a {CI} rather than as {WCs} or {MH} insulators, except at $ \nu$=1 which is a strict MI. {Quite generally, WC and MI are adiabatically connected—with the WC being the weak lattice potential and low-carrier-density (and consequently, the vanishing filling factor) limit of the {CI} whereas the MI is the strong lattice potential (and hence flatband) half-filling limit of the {CI} (in specific situations)~\cite{vu2020collective}.}
For the half-filled $\nu$=1 situation, the MI description is the appropriate description because this insulator arises at half-filling independent of how high the carrier density might be. {Additionally}, as shown in Fig.~\ref{fig:fig1}, the $\nu$=1 {MI} exists independent of whether the interaction is {long- or short-ranged}.  But the insulator at other fillings (e.g., $\nu$=3/4, 2/3, 1/2, 1/3, 1/4 as discussed in Ref.~\onlinecite{pan2020quantum}) cannot be a simple MI since they {disappear} for the on-site interaction-only model. We have explicitly checked that the insulating phase at {all $\nu$ except for $ \nu $=1} disappears if the distant neighbor interaction terms are {zero} in Eq.~\eqref{eq:hubbard}.  The fact that the existence of {CIs} at $\nu$ {other than 1} depends crucially on having a long-range interaction may indicate that the WC terminology is more appropriate for the {CI} for $\nu$ other than unity.  But this is {untrue} as can be seen from Fig.~\ref{fig:fig4} where we show the dimensionless continuum Coulomb coupling $r_s$, which is the average interparticle separation measured in units of the effective {Bohr radius as a function} of $\epsilon$ and $\nu$ for a {fixed} $\theta$=4 in Fig.~\ref{fig:fig4}(a) and as a function of $ \theta $ and $ \nu $ for a fixed $ \epsilon $=5 in Fig.~\ref{fig:fig4}(b).  {It is obvious} that the applicable $r_s$ {for TTMD are too small} in the physical parameter regime (i.e., $\epsilon>$5 and $\nu>$1/4) for the WC to occur as it is well-established that the critical $r_s$ necessary for a 2D WC is $ r_s> $ 30~\cite{drummond2009phase}. {These} {CIs} at simple rational fillings other than 1 are better {considered} as the lattice versions of a WC [i.e., quantum charge density wave ordering (QCDW)]. We predict that there should be a similar interaction-induced {MI} {in TBLG at $ \nu $= 1/8}, etc. ({like} $ \nu $=1/2, etc. in TTMD), where electrons occupy distant TBLG moir\'e unit cells forming a {QCDW}.

Our finding that a TTMD half-filled doped hole system is an antiferromagnetic {MI} is {significant} since the corresponding TBLG {CI} ground states are {deemed} ferromagnetic Chern insulators~\cite{sharpe2019emergent,serlin2020intrinsic,wu2020collective,wu2020quantum,bultinck2020mechanism,alavirad2020ferromagnetism}. Establishing the antiferromagnetic spin configuration of a TTMD half-filled moir\'e system is an important future experimental challenge.

We have theoretically discussed the filling-factor tuned {MIT} in 2D moir\'e TTMD hole doped materials {as well as the nature of the correlated insulating states at rational fillings (which are all nominal metals in the noninteracting band pictures)}. We establish the existence of an antiferromagnetic Mott insulator at half-filling and the emergence of nearby (i.e., just away from half-filling) magnetic metallic phases. Our results are reminiscent of the Nagaoka ferromagnetism inherent in the strongly interacting Hubbard model around half-filling although we consider finite (albeit low) doping level~\cite{nagaoka1966ferromagnetism},{ and are consistent with recent experiments reporting metallicity in TTMD slightly away from half-filling~\cite{ghiotto2021quantum,li2021continuous}. We predict such metal-insulator transitions at other rational fillings such as $\nu=1/3$ ~\cite{MI_SM}.}

We gratefully thank Cory Dean and Abhay Pasupathy (and their team members, {especially Augusto Ghiotto}) for discussions on their unpublished experimental results, {and Fengcheng Wu for helpful discussion}.  This work is supported by the Laboratory for Physical Sciences.

\bibliographystyle{apsrev4-1}
\bibliography{M-I}

\end{document}